\documentclass[aps,prl,twocolumn,showpacs,superscriptaddress,groupedaddress,acknowledgment]{revtex4-1}  
\usepackage{dcolumn}   
\usepackage{bm}        

\usepackage{amsmath,comment}
\usepackage{braket}
\usepackage{amsfonts}
\usepackage{graphicx}
\usepackage{amsthm}
\usepackage{color}
\usepackage{setspace}
\usepackage{amssymb}
\usepackage{latexsym}
\usepackage{here}

\theoremstyle{definition}

\newtheorem*{theorem*}{Theorem}

\newtheorem*{definition*}{Definition}

\newcommand{\mc}[1]{\mathcal{#1}}
\newcommand{\mr}[1]{\mathrm{#1}}
\newcommand{\mbb}[1]{\mathbb{#1}}

\newcommand{\lrs}[1]{\left( #1 \right)}
\newcommand{\lrm}[1]{{\left\{ #1 \right\}}}
\newcommand{\lrl}[1]{\left[ #1 \right]}

\newcommand{\fracpd}[2]{\frac{\partial #1 }{\partial #2 }}

\newcommand{\aln}[1]{
\begin{align}
#1
\end{align}
}

\newcommand{\ra}{\rightarrow}

\newcommand{\Tr}{\mr{Tr}}

\begin{document}
\title{
Operator Noncommutativity and Irreversibility in Quantum Chaos
}
\date{\today}
\author{Ryusuke Hamazaki$^1$, Kazuya Fujimoto$^1$, and Masahito Ueda$^{1,2}$}
\affiliation{
$^1$Department of Physics, University of Tokyo, 7-3-1 Hongo, Bunkyo-ku, Tokyo 113-0033, Japan\\
$^2$RIKEN Center for Emergent Matter Science (CEMS), Wako 351-0198, Japan
}

\begin{abstract}
We argue that two distinct probes of quantum chaos, i.e.,
the growth of noncommutativity of two unequal-time operators and the degree of irreversibility  in a time-reversal test, are equivalent for initially localized states.
We confirm this for interacting nonintegrable many-body systems and a quantum kicked rotor.
Our results show that three-point out-of-time-ordered correlators dominate the growth of the squared commutator for initially localized states, in stark contrast to four-point out-of-time-ordered correlators that have extensively been studied for thermal initial states.

\end{abstract}
\pacs{05.30.-d, 05.45.Mt}

\maketitle
\paragraph{Introduction.}
Quantum chaos~\cite{Haake,Stockmann00,Gutzwiller13} has attracted considerable interest since the late 1970s~\cite{Berry77R,Peres84S,Bohigas84,Berry89,Andreev96}.
While the eigenvalue/vector statistics of a Hamiltonian have often been studied as probes of quantum chaos~\cite{Berry77R,Peres84E2,Bohigas84,Srednicki94,Muller04,Rigol08,Santos10a,Pal10,Khatami13,Beugeling14,Beugeling15,Mondaini16,Mondaini17,DAlessio16,Luitz16,Serbyn16S,Hamazaki18A,Kos17}, it can also be characterized by its dynamics as in classical chaos~\cite{Strogatz18}.
A prime indicator of such dynamical characterization is irreversibility~\cite{Thomson74} in chaotic motion.
A prototypical examples is a time-reversal test~\cite{Adachi88,Yamada12,Murashita18} in which a system evolves forward and then backward in time for the same time period by adding a small perturbation in the return process.
If the dynamics is chaotic, the final state deviates significantly from the initial state however small the perturbation is~\cite{Shepelyansky83,Murashita18}.
References ~\cite{Adachi88,Yamada12,Schmitt17,Schmitt18} discuss the irreversibility in quantum chaos measured by expectation values of observables under the time-reversal test with a unitary perturbation added upon reversal~\cite{Los-OTOC}.
Note that localized initial states are suitable for the study of irreversible delocalization of the state under the time-reversal test~\cite{Shepelyansky83,Peres84S,Adachi88,Jalabert01,Yamada12,Schmitt17,Schmitt18}.

As another dynamical probe of quantum chaos, the growth of quantum noncommutativity of two unequal-time operators has recently been proposed independently of irreversibility.
In particular, an expectation value of a squared commutator of two unequal-time observables  has actively been investigated in various fields ranging from high-energy~\cite{Kitaev14H,Kitaev15S,Shenker15S,Maldacena16,Maldacena16R,Polchinski16S} to condensed-matter physics~\cite{Gu16,Bohrdt16,Garttner17,Banerjee17,Shen17,Fan17,Huang17O,Li17,Tsuji17,Kukuljan17,Dora17,Iyoda16,Patel17,Jian18}.
For semiclassical chaotic models before the Ehrenfest time~\cite{Larkin69,Kitaev15S,Kurchan16,Rozenbaum17,Cotler17,Hashimoto17,Scaffidi17,Das17,Rozenbaum18}, the semiclassical approximation ensures that the squared commutator grows exponentially reflecting the instability of phase-space trajectories.

Despite the surge of interest, how the squared commutator is related to other conventional chaotic probes has remained elusive~\cite{Iyoda16,Rozenbaum17,Rozenbaum18}, and its relevance to  irreversibility is an intriguing problem.
Several studies indeed suggested the qualitative similarity between the squared commutator and the Loschmidt echo~\cite{Hashimoto17,Rozenbaum17,Garttner18,Tsuji18}, but quantitative understanding has remained an open issue.
Another study~\cite{Schmitt18} reports that a certain type of commutators appears in the expansion of the irreversibility measure after the time-reversal test~\cite{Sch-OTOC}.
However, it is unclear how their results are related to  previous discussions on noncommutativity growth based on the squared commutator.

\begin{figure}[t]
\begin{center}
\includegraphics[width=\linewidth]{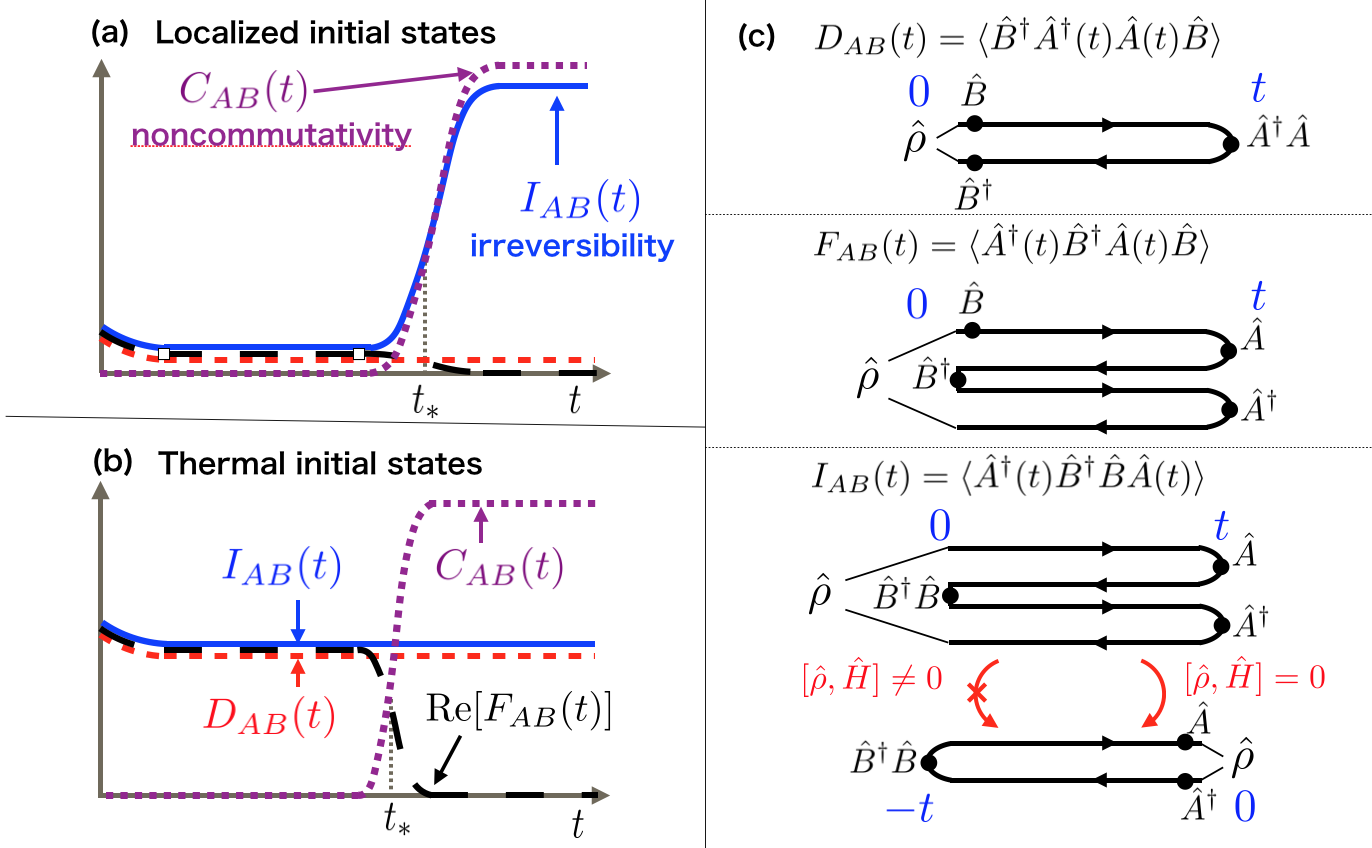}
\caption{
(a) Schematic behaviors of the unequal-time correlators and the squared commutator $C_{AB}(t)=\braket{|[\hat{A}(t),\hat{B}]|^2}$ (dotted) for initially localized states. Irreversibility $I_{AB}(t)=\braket{\hat{A}(t)^\dag\hat{B}^\dag\hat{B}\hat{A}(t)}$ (solid) dominates the growth of noncommutativity $C_{AB}(t)$.
(b) For initially thermal states, only $F_{AB}(t)=\braket{\hat{A}(t)^\dag\hat{B}^\dag\hat{A}(t)\hat{B}}$ (long-dashed) is considered to show a nontrivial decay around a timescale $t_*$, contributing to $C_{AB}(t)$, while $I_{AB}(t)$ and $D_{AB}(t)=\braket{\hat{B}^\dag\hat{A}(t)^\dag\hat{A}(t)\hat{B}}$ (short-dashed) do not show nontrivial behavior.
(c) The three-point function $D_{AB}(t)$ is time-ordered and the four-point function $F_{AB}(t)$ is out-of-time-ordered.
The three-point correlator $I_{AB}(t)$ becomes anti-time-ordered if $[\hat{\rho},\hat{H}]=0$ because it is then equal to $\langle \hat{A}^\dagger\hat{B}^\dagger(-t)\hat{B}(-t)\hat{A}\rangle$.
Such a reduction does not occur if $[\hat{\rho},\hat{H}]\neq 0$.
}
\label{figure1}
\end{center}
\end{figure}

In this Letter, we argue that noncommutativity and irreversibility are essentially equivalent to each other for initially localized states.
Namely, the squared commutator $C_{AB}(t):=\braket{|[\hat{A}(t),\hat{B}]|^2}$ of two unequal-time observables $\hat{A}(t)$ and $\hat{B}=\hat{B}(0)$ is equivalent to $I_{AB}(t):=\braket{\hat{A}(t)^\dag\hat{B}^\dag\hat{B}\hat{A}(t)}$, which is interpreted as the irreversibility  measured through $\hat{B}$ (we assume $\braket{\hat{B}}=0$) under the time-reversal test against perturbation $\hat{A}$ at time $t$.
In fact, we prove the following relation:
\aln{\label{main}
\frac{C_{AB}(t)}{I_{AB}(t)}=\lrs{1+\alpha_t\sqrt{\frac{D_{AB}(t)}{I_{AB}(t)}}}^2,
}
where
$\alpha_t$ is a time-dependent numerical factor satisfying $|\alpha_t|\leq 1$,
$D_{AB}(t):=\braket{\hat{B}^\dag\hat{A}(t)^\dag\hat{A}(t)\hat{B}}$ is a time-ordered correlator.
We show $\frac{D_{AB}(t)}{I_{AB}(t)}\ll 1$ and hence $C_{AB}(t)\simeq I_{AB}(t)$ for initially localized states in chaotic systems (see Fig.~\ref{figure1}(a)), by decomposing $\frac{D_{AB}(t)}{I_{AB}(t)}$ into physical processes that involve the time-reversal test (see Eqs.~(\ref{juyou})-(\ref{jo1})).
We do not intend to investigate the existence of irreversibility in quantum systems by using noncommutativity.
Our aim is to find the nontrivial relation between the two distinct probes of chaos, by noticing the importance of initially localized states.

Our work is fundamentally important especially in the rapidly growing community of  out-of-time-ordered correlators (OTOC)~\cite{Cor-OTOC}.
The squared commutator can be decomposed as $C_{AB}(t)=I_{AB}(t)+D_{AB}(t)-2\mr{Re}[F_{AB}(t)]$,
where $F_{AB}(t):=\braket{\hat{A}(t)^\dag\hat{B}^\dag\hat{A}(t)\hat{B}}$ is an OTOC (we do not call $C_{AB}(t)$ the OTOC here).
Previous studies mainly considered delocalized thermal-equilibrium initial states and argued that the dynamics of four-point OTOC (4-OTOC) $F_{AB}(t)$ contributes to a nontrivial growth of $C_{AB}(t)$ around a timescale $t_*$, while $I_{AB}(t)$ and $D_{AB}(t)$ rapidly decay to constant values much before $t_*$ (Fig.~\ref{figure1}(b))~\cite{Kitaev14H,Kitaev15S,Hosur16,Maldacena16R,Huang17O}.
In this case, $[\hat{H},\hat{\rho}]=0$ and thus
$I_{AB}(t)$ is (anti-)time-ordered (Fig.~\ref{figure1}(c)).
For nonequilibrium states satisfying $[\hat{H},\hat{\rho}]\neq 0$, we find that
the three-point correlator $I_{AB}(t)$ also becomes an OTOC, which we refer to as a three-point OTOC (3-OTOC).
From $C_{AB}(t)\simeq I_{AB}(t)$, we argue that the 3-OTOC $I_{AB}(t)$ rather than the 4-OTOC $F_{AB}(t)$ dominates the growth of $C_{AB}(t)$ for initially localized states (Fig.~\ref{figure1}(a))~\cite{Change-OTOC}.
The importance of the 3-OTOC for noncommutativity has never been reported before.

\begin{figure}[t]
\begin{center}
\includegraphics[width=8cm]{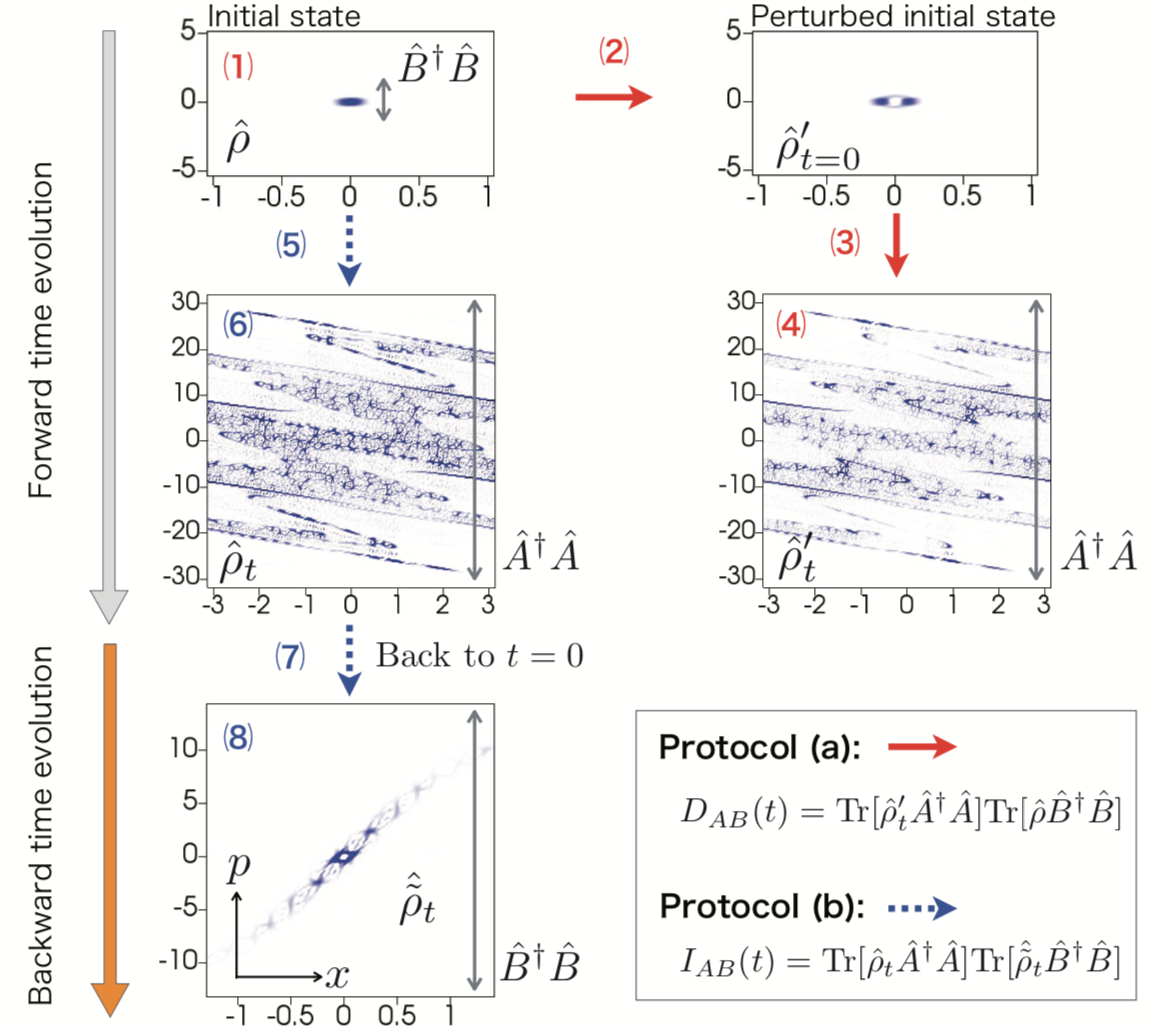}
\caption{
Coarse-grained Wigner functions obtained from an initially localized state $\hat{\rho}=\hat{\rho}_\mr{w}$ with width $\sigma=1$, the strength of the kick $K=4$ and an effective Planck constant $\hbar_\mr{eff}=2^{-7}$ for the quantum kicked rotor in Eq.~(\ref{krotor}).
Protocol (a) (red solid arrows) consists of (1) obtaining the expectation value of $\hat{B}^\dag\hat{B}$ for the initial state $\hat{\rho}$, (2) perturbation by $\hat{B}$ as $\hat{\rho}'=\frac{\hat{B}\hat{\rho}\hat{B}^\dag}{\Tr[\hat{\rho}\hat{B}^\dag\hat{B}]}$, (3) unitary time evolution $\hat{U}_t$, giving $\hat{\rho}_t'=\hat{U}_t\hat{\rho}'\hat{U}_t^\dag$, and (4) obtaining the expectation value of $\hat{A}^\dag\hat{A}$. The multiplication of the two expectation values gives $D(t)=\braket{\hat{B}^\dag\hat{A}^\dag(t)\hat{A}(t)\hat{B}}$.
Protocol (b) (blue dashed arrows) consists of (5) unitary time evolution $\hat{U}_t$, giving $\hat{\rho}_t$, (6) obtaining the expectation value of $\hat{A}^\dag\hat{A}$,
(7) perturbation by $\hat{A}$ as $\frac{\hat{A}\hat{\rho}_t\hat{A}^\dag}{\Tr[{\hat{\rho}_t\hat{A}^\dag\hat{A}}]}$ and backward time evolution $\hat{U}_t^\dag$, giving $\hat{\widetilde{\rho}}_t =\frac{\hat{A}(t)\hat{\rho}\hat{A}(t)^\dag}{\Tr[\hat{\rho}\hat{A}(t)^\dag\hat{A}(t)]}$, and
(8) obtaining the expectation value of $\hat{B}^\dag\hat{B}$.
 The multiplication of the two expectation values gives $I(t)=\braket{\hat{A}^\dag(t)\hat{B}^\dag\hat{B}\hat{A}(t)}$.
For these Wigner functions, the initial state $\hat{\rho}=\hat{\rho}_\mr{w}$, which is localized around $(x,p)=(0,0)$, spreads in the course of time evolution.
The amount of spreading of $\hat{\rho}_{t}$ at time $t=6$ is similar to that of $\hat{\rho}'_{t}$,
which leads to (i) in Eq.~(\ref{jo1}) with $\hat{A}=\hat{p}$ (see also Eq.~(\ref{teigi})).
On the other hand, $\hat{\widetilde{\rho}}_{t}$ spreads rapidly,
leading to (ii) in Eq.~(\ref{jo1}) with $\hat{B}=\hat{p}$.
}
\label{modoru2}
\end{center}
\end{figure}

\paragraph{Irreversibility and noncommutativity.}
We first outline the proof of Eq.~(\ref{main}) (see Appendix I in detail~\cite{footOTOC1}).
The Cauchy-Schwarz inequality leads to $|F_{AB}(t)|\leq \sqrt{I_{AB}(t)D_{AB}(t)}$.
Then, we find $(C_{AB}(t)-I_{AB}(t)-D_{AB}(t))^2\leq 4I_{AB}(t)D_{AB}(t)$.
The positivity of $C_{AB}(t),I_{AB}(t),$ and $D_{AB}(t)$ leads to $|\sqrt{C_{AB}(t)}-\sqrt{I_{AB}(t)}|\leq \sqrt{D_{AB}(t)}$.
Dividing both sides by $I_{AB}(t)$ and introducing $|\alpha_t|\leq 1$, we obtain Eq.~(\ref{main}).

We now show how $I_{AB}(t)$ measures the system's irreversibility for an initially localized state.
We define localized initial states with respect to $\hat{B}$ as states that satisfy $\braket{\hat{B}^\dag\hat{B}}\ll\braket{\hat{B}^\dag(t)\hat{B}(t)}$ for sufficiently large $t$.
We first decompose $D_{AB}(t)$ and $I_{AB}(t)$ as
\aln{\label{juyou}
D_{AB}(t)=\braket{\hat{B}^\dag\hat{A}(t)^\dag\hat{A}(t)\hat{B}} &=\Tr[{\hat{\rho}_t'\hat{A}^\dag\hat{A}}]\Tr[{\hat{\rho}\hat{B}^\dag\hat{B}}],\nonumber\\
I_{AB}(t)=\braket{\hat{A}(t)^\dag\hat{B}^\dag\hat{B}\hat{A}(t)} &=\Tr[{\hat{\rho}_t\hat{A}^\dag\hat{A}}]\Tr[{\hat{\widetilde{\rho}}_t\hat{B}^\dag\hat{B}}],
}
where $\hat{\rho}_t:=\hat{U}_t\hat{\rho}\hat{U}_t^\dag$, $\hat{\rho}_t':=\frac{\hat{U}_t\hat{B}\hat{\rho}\hat{B}^\dag \hat{U}_t^\dag}{\Tr[{\hat{\rho}\hat{B}^\dag\hat{B}}]}$, and $\hat{\widetilde{\rho}}_t:=\frac{\hat{A}(t)\hat{\rho}\hat{A}(t)^\dag}{\Tr[{\hat{\rho}\hat{A}(t)^\dag\hat{A}(t)}]}$ are the states obtained from the initial state $\hat{\rho}$ by two protocols (a) and (b) as illustrated in Fig.~\ref{modoru2}.
This figure shows the coarse-grained Wigner functions on the $x$-$p$ phase space for each density matrix, which is numerically obtained by using the quantum kicked rotor.
Here, we focus on how they change for two different protocols.
In protocol (a), the initial state is perturbed with $\hat{B}$ as $\hat{\rho}'=\frac{\hat{B}\hat{\rho}\hat{B}^\dag}{\Tr[{\hat{\rho}\hat{B}^\dag\hat{B}}]}$, which then evolves in time as
$\hat{\rho}_t':=\hat{U}_t\hat{\rho}'\hat{U}_t^\dag$.
The product of the expectation value of $\hat{B}^\dag\hat{B}$ for $\hat{\rho}$ and that of $\hat{A}^\dag\hat{A}$ for $\hat{\rho}_t'$ gives $D_{AB}(t)$.
For protocol (b), we let the state evolve during time $t$ as $\hat{\rho}_t=\hat{U}_t\hat{\rho}\hat{U}_t^\dag$, and  perturb the state with $\hat{A}$ as $\frac{\hat{A}\hat{\rho}_t\hat{A}^\dag}{\Tr[{\hat{\rho}_t\hat{A}^\dag\hat{A}}]}$.
We then perform time reversal $t\ra -t$, obtaining $\hat{\widetilde{\rho}}_t=\frac{\hat{U}_t^\dag\hat{A}\hat{U}_t\hat{\rho}\hat{U}_t^\dag\hat{A}^\dag\hat{U}_t}{\Tr[{\hat{U}_t\hat{\rho}\hat{U}_t^\dag\hat{A}^\dag\hat{A}}]}=\frac{\hat{A}(t)\hat{\rho}\hat{A}(t)^\dag}{\Tr[{\hat{\rho}\hat{A}(t)^\dag\hat{A}(t)}]}$~\cite{Pert-OTOC}.
The product of the expectation value of $\hat{B}^\dag\hat{B}$ for $\hat{\widetilde{\rho}}_t$ and that of $\hat{A}^\dag\hat{A}$ for $\hat{\rho}_t$ gives $I_{AB}(t)$.
Thus, $I_{AB}(t)$ involves the time-reversal test and measures the degree of irreversibility.
Note that this protocol is similar to the one used in Refs.~\cite{Adachi88,Yamada12,Schmitt17,Schmitt18}, where $\hat{A}$ is chosen to be  unitary.

From Eq.~(\ref{juyou}), we obtain $\frac{D_{AB}(t)}{I_{AB}(t)}=s_{t}r_{t}$, where
\aln{\label{teigi}
s_{t}:=\frac{\Tr[\hat{\rho}_t'\hat{A}^\dag\hat{A}]}{\Tr[\hat{\rho}_t\hat{A}^\dag\hat{A}]},\:\:\:\:r_{t}:=\frac{\Tr[\hat{\rho}\hat{B}^\dag\hat{B}]}{\Tr[\hat{\widetilde{\rho}}_t\hat{B}^\dag\hat{B}]}.
}
Equation~(\ref{main}) then becomes
\aln{\label{atarasi}
\frac{C_{AB}(t)}{I_{AB}(t)}=\lrs{1+\alpha_t\sqrt{s_tr_t}}^2.
}

Now, let us consider two conditions
\aln{
\label{jo1}
(\mr{i})\:\:s_{t}= \mr{O}(1),\:\:\:\:(\mr{ii})\:\:r_{t}\ll 1.
}
The condition (i) means that time evolution of $\hat{A}^\dag\hat{A}$ is stable under the initial perturbation of $\hat{B}$.
The condition (ii) means that the initial state $\hat{{\rho}}$ is irreversible due to the sensitivity against the time-reversal test.
Note that this is macroscopic irreversibility, where the irreversibility appears not only in the density matrix but also in the expectation value of the macroscopic observable $\hat{B}^\dag\hat{B}$~\cite{Macro-OTOC}.
We argue that these two conditions hold true for a wide class of chaotic dynamics with initially localized states, and hence $C_{AB}(t)\simeq I_{AB}(t)$.
In the following, we test this conjecture for  quantum many-body systems and a kicked rotor.
For the former systems, we also argue that the condition (ii) breaks down for initially thermal states under certain assumptions.

\paragraph{Interacting quantum many-body systems.}
We first consider locally interacting many-body systems on $N$ lattice sites.
While we do not have a simple phase-space representation as in Fig.~\ref{modoru2}, the above protocols are well-defined.
Note that our system is without a well-defined semiclassical limit and exponential sensitivity against initial perturbation.
We especially focus on $\hat{H}$ and $\hat{B}$ that can be written as $\sum_{i}\hat{h}_i$ and $\sum_{i}\hat{b}_i$, where translationally invariant operators $\hat{h}_i$ and $\hat{b}_i$ are independent of the site $i$.
We also assume $\braket{\hat{B}}=0$, which leads to $\braket{\hat{b}_i}=0$.
We consider a translationally invariant initial state $\hat{\rho}$ that satisfies the cluster decomposition property~\cite{Ruelle99},
which means that for two distant regions $\mc{I}$ and $\mc{J}$ with $\mc{I}\cap\mc{J}=\varnothing$, $\braket{\prod_\lrm{f|k_f\in\mc{I}\cup\mc{J}}\hat{a}_{k_f}^f}\simeq \braket{\prod_\lrm{f|k_f\in\mc{I}}\hat{a}_{k_f}^f}\braket{\prod_\lrm{f|k_f\in\mc{J}}\hat{a}_{k_f}^f}$.
Here $\hat{a}_{k_f}^f$ is the $f$-th operator localized around site $k_f$.
From the cluster decomposition, we can show that the energies of $\hat{\rho}$ and $\hat{\rho}'$ are macroscopically equal (see Appendix IIA~\cite{footOTOC1} for a proof).

To justify condition (i) in Eq.~(\ref{jo1}), we invoke the eigenstate thermalization hypothesis (ETH)~\cite{Deutsch91,Srednicki94,Rigol08}, which is expected to hold for nonintegrable systems~\cite{ETH-OTOC}.
The ETH justifies that any initial state with a given energy relaxes to a state described by the canonical ensemble at the corresponding temperature for most of the time~\cite{Srednicki94,Rigol08}.
By applying the ETH, we find that $\hat{\rho}$ and $\hat{\rho}'$  relax to the same canonical ensemble at inverse temperature $\beta$ in the long run because they have the same energies.
Thus, $s_{t}\simeq 1$ for most of the time, which justifies the condition (i) in Eq.~(\ref{jo1}).

To find a sufficient condition for (ii) in Eq.~(\ref{jo1}), we additionally assume that
$\hat{A}$ can be written as a local operator or the sum of local operators.
Then, using the ETH, we obtain $r_t^{-1}\simeq \frac{|\braket{\hat{A}}_\beta|^2}{\braket{\hat{A}^\dag\hat{A}}_\beta}+\lrs{1-\frac{|\braket{\hat{A}}_\beta|^2}{\braket{\hat{A}^\dag\hat{A}}_\beta}}\frac{\braket{\hat{B}^\dag\hat{B}}_\beta}{\braket{\hat{B}^\dag\hat{B}}}$ (see Appendix IIB~\cite{footOTOC1}).
For initially localized states, $\Tr[\hat{\rho}\hat{B}^\dag\hat{B}]=\braket{\hat{B}^\dag\hat{B}}\ll\braket{\hat{B}^\dag(t)\hat{B}(t)}\simeq \braket{\hat{B}^\dag\hat{B}}_{\beta}$ because of their definition and the ETH,
and thus $r_t\ll 1$ in the thermodynamic limit if $1-\frac{|\braket{\hat{A}}_\beta|^2}{\braket{\hat{A}^\dag\hat{A}}_\beta}$ is $\mr{O}(N^0)$.
As discussed in Appendix IIB~\cite{footOTOC1}, this condition is satisfied e.g., if $\hat{A}$ is an observable with $\braket{\hat{A}}_\beta=0$~\cite{Change-OTOC} and is a typical local operator.
On the other hand, for initially thermal states (or other delocalized states that are equivalent to thermal states for macroscopic observables~\cite{Goldstein06,Popescu06}), $r_t\simeq 1$ because $\braket{\hat{B}^\dag\hat{B}}=\braket{\hat{B}^\dag\hat{B}}_\beta$ and the condition (\ref{jo1}) (ii) does not hold.

The initially localized state defined above, which satisfies $\braket{\hat{B}^\dag\hat{B}}\ll \braket{\hat{B}^\dag\hat{B}}_{\beta}$, is naturally obtained for nonequilibrium states.
To see this, we assume that the canonical ensemble has the cluster decomposition property,
which is satisfied for sufficiently high-temperature systems in our setup~\cite{Kliesch15}.
We also assume that $\braket{\hat{b}_i}_\beta$ is nonzero in the thermodynamic limit.
Then, we obtain $\braket{\hat{B}^\dag\hat{B}}\simeq  \sum_{(i,j)\in\mc{A}_0}\braket{\hat{b}_i^\dag\hat{b}_j}=\mr{O}(N)$ and $\braket{\hat{B}^\dag\hat{B}}_{\beta}\simeq \sum_{(i,j)\in\mc{A}_0}\braket{\hat{b}_i^\dag\hat{b}_j}_{\beta}+\sum_{(i,j)\notin\mc{A}_0}\braket{\hat{b}_i^\dag}_{\beta}\braket{\hat{b}_j}_{\beta}=\mr{O}(N^2)$.
Here, $\mc{A}_0$ is a set where $i$ and $j$ are close.
Thus, our assumptions above lead to $\braket{\hat{B}^\dag\hat{B}}\ll \braket{\hat{B}^\dag\hat{B}}_{\beta}$ in the thermodynamic limit.
From this discussion, we also obtain $r_{t}=\mr{O}(N^{-1})$.

\begin{figure}[t]
\begin{center}
\includegraphics[width=\linewidth]{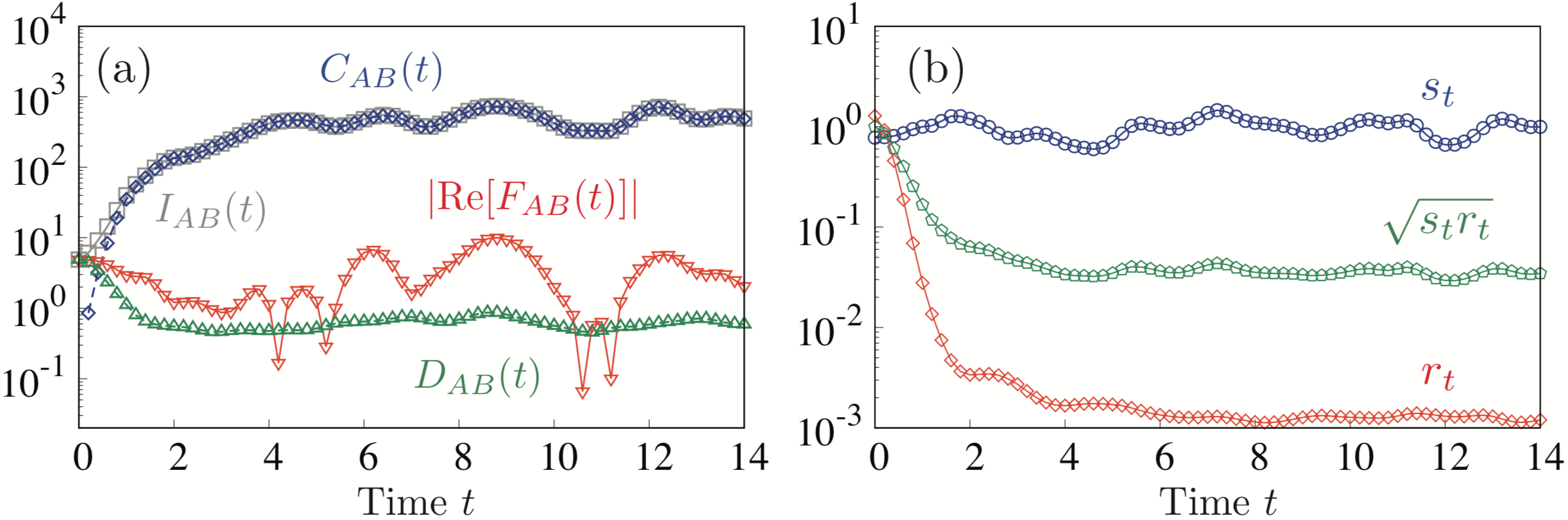}
\caption{Quench dynamics of the transverse Ising model $H_\mr{TI}(h=0.5)$ for $L=14$.
Here, $\hat{A}=\sum_{i=1}^L(\hat{\sigma}_i^z-\braket{\hat{\sigma}_i^z}_\beta)$ ($\beta$ is determined from the total energy after quench) and $\hat{B}=\sum_{i=1}^L(\hat{\sigma}_i^z-\braket{\hat{\sigma}_i^z})$, and the initial state is the ground state of $\hat{H}_\mr{TI}(h=-5.0)$.
The Planck constant is set to be unity.
(a) Time evolutions of $C_{AB}(t)=-\braket{[\hat{A}(t),\hat{B}]^2},$ $I_{AB}(t)=\braket{\hat{A}(t)^\dag\hat{B}^\dag\hat{B}\hat{A}(t)},$ $D_{AB}(t)=\braket{\hat{B}^\dag\hat{A}(t)^\dag\hat{A}(t)\hat{B}}$ and $|\mr{Re}[F_{AB}(t)]|=|\mr{Re}[\braket{\hat{A}(t)^\dag\hat{B}^\dag\hat{A}(t)\hat{B}}]|$.
The approximated equality $C_{AB}(t)\simeq I_{AB}(t)$ holds true for $t\gtrsim 1$.
(b) Time evolutions of $s_{t},r_{t}$ and $\sqrt{s_{t}r_{t}}$ defined in Eq.~(\ref{teigi}).
When $t\gtrsim 1$, two conditions (\ref{jo1}) are satisfied and the second term in Eq.~(\ref{atarasi}) becomes small.
The curves are obtained by using the QUSPIN package~\cite{Weinberg17}.
}
\label{mb}
\end{center}
\end{figure}

We numerically check Eq.~(\ref{jo1}) and $C_{AB}(t)\simeq I_{AB}(t)$ for a 1D transverse Ising model after a sudden quench.
The Hamiltonian is given by~\cite{Kim15} $\hat{H}_\mr{TI}(h):=-\sum_{i=1}^L\hat{\sigma}_i^z\hat{\sigma}_{i+1}^z-1.05\hat{\sigma}_i^x+h\hat{\sigma}_i^z$ with a periodic boundary $\hat{\sigma}_{L+1}^z=\hat{\sigma}_1^z$.
As an initial state, we consider the ground state of $\hat{H}_\mr{TI}(h)$ with $h=-5$,
which is close to the state where all spins are polarized upwards, i.e., localized in this direction.
Then, we suddenly changes the value of $h$ to $h=0.5$.
Figure~\ref{mb} (a) shows time evolutions of $C_{AB}(t), I_{AB}(t), D_{AB}(t)$ and $|\mr{Re}[F_{AB}(t)]|$ with $\hat{A}=\sum_{i=1}^L(\hat{\sigma}_i^z-\braket{\hat{\sigma}_i^z}_\beta)$ ($\beta$ is determined from the total energy after the quench) and $\hat{B}=\sum_{i=1}^L(\hat{\sigma}_i^z-\braket{\hat{\sigma}_i^z})$ for $L=14$.
We see that $I_{AB}(t)$ and $C_{AB}(t)$ behave almost identically, while the other functions do not grow much.
Figure~\ref{mb} (b) shows time evolutions of $s_{t},r_{t}$ and $\sqrt{s_{t}r_{t}}$.
We see that two conditions (\ref{jo1}) are satisfied when $t\gtrsim 1$ (namely, even before the dynamics becomes stationary), and that the second term in Eq.~(\ref{atarasi}) becomes small.

\paragraph{Quantum kicked rotor.}
Next, we numerically confirm Eq.~(\ref{jo1}) and $C_{AB}(t)\simeq I_{AB}(t)$ for a single-particle quantum kicked rotor:
\aln{\label{krotor}
\hat{H}_\mr{QKR}(t):=\frac{\hat{p}^2}{2}+K\cos \hat{x}\sum_{n}\delta(t-n),
}
where $\hat{p}=-i\hbar_\mr{eff}\fracpd{}{x}$ is the (angular) momentum operator and $\hbar_\mr{eff}$ denotes the dimensionless Planck constant, which scales with $\hbar$.
We impose a periodic boundary condition on  $x$ as $-\pi\leq x<\pi$.
Then, $\hat{p}$ has eigenvalues $m\hbar_\mr{eff}$ and eigenvectors $\braket{x|p_m} =\frac{1}{\sqrt{2\pi}}e^{imx}$ for each $m\:(m\in\mbb{Z})$.
We consider an initial wave-packet state
$
\hat{\rho}_\mr{w}:=\ket{\psi_\mr{w}}\bra{\psi_\mr{w}},\: \ket{\psi_\mr{w}}:=\frac{1}{Z_\mr{w}}\sum_m e^{-\frac{\hbar_\mr{eff} m^2}{2\sigma^2}}\ket{p_m},
$
where $Z_\mr{w}:=\sqrt{\sum_m e^{-\frac{\hbar_\mr{eff}m^2}{\sigma^2}}}$.
See Appendix IIIA in Supplemental Material~\cite{footOTOC1} for different initial states.

\begin{figure}
\begin{center}
\includegraphics[width=\linewidth]{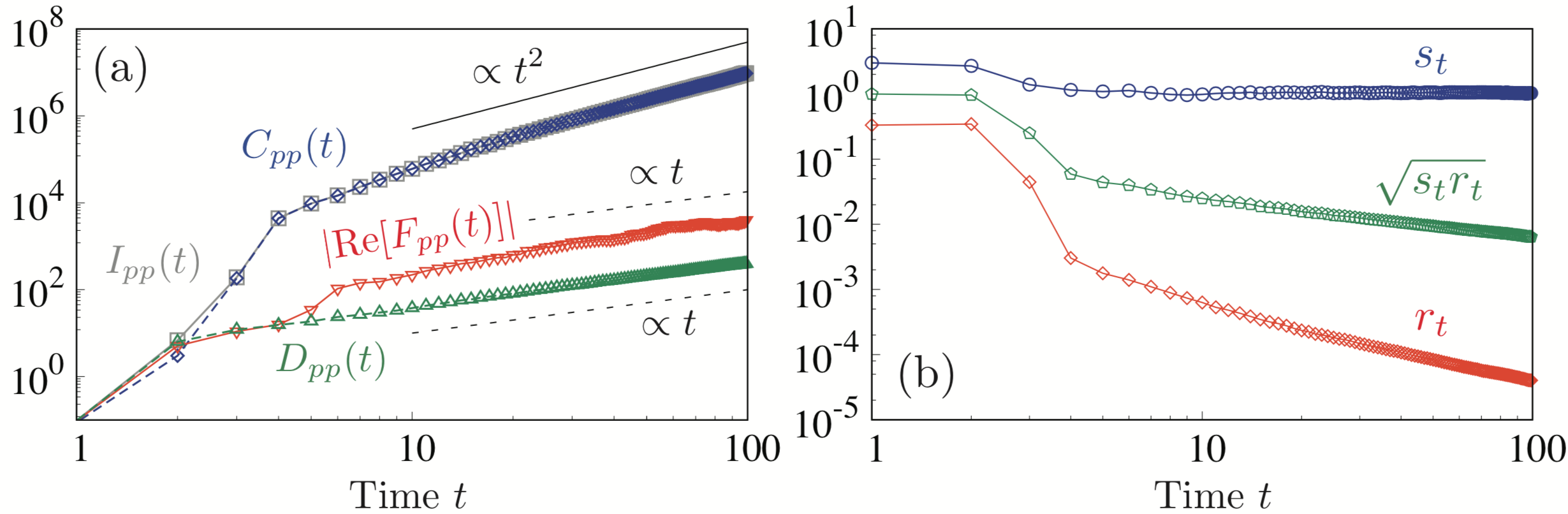}
\caption{
Dynamics of the quantum kicked rotor $\hat{H}_\mr{QKR}(t)$ with $\hbar_\mr{eff}=2^{-6},\sigma=4$, and $K=10$.
We take $\hat{A}=\hat{B}=\hat{p}$ and the initial wave-packet state $\hat{\rho}_\mr{w}$.
(a) Time evolutions of $D_{pp}(t)=\braket{\hat{p}\hat{p}(t)^2\hat{p}}$, $|\mr{Re}[F_{pp}(t)]|=|\mr{Re}[\braket{\hat{p}(t)\hat{p}\hat{p}(t)\hat{p}}]|$, $I_{pp}(t)=\braket{\hat{p}(t)\hat{p}^2\hat{p}(t)}$, and $C_{pp}(t)=-\braket{[\hat{p}(t),\hat{p}]^2}$.
Two dashed lines and a solid line show the linear and quadratic dependences on $t$, respectively.
 For $t\gtrsim 3$, $C_{pp}(t)\simeq I_{pp}(t)$ holds true.
(b) Time evolutions of $s_{t},r_{t}$ and $\sqrt{s_{t}r_{t}}$ defined in Eq.~(\ref{teigi}).
When $t\gtrsim 3$, two conditions (\ref{jo1}) are satisfied and the second term in Eq.~(\ref{atarasi}) becomes small.
}
\label{qkr}
\end{center}
\end{figure}

We consider $\hat{A}=\hat{B}=\hat{p}$ after $t\:(\in\mbb{Z})$ periods.
Note that $\hat{\rho}_\mr{w}$ is initially localized with respect to $\hat{p}$ because of $\braket{\hat{p}}=0$ and $\braket{\hat{p}^2}\ll \braket{\hat{p}(t)^2}\:(\propto t)$~\cite{Chirikov88} for large $t$.
Figure~\ref{qkr} (a) shows the dynamics of $C_{pp}(t)=-\braket{[\hat{p}(t),\hat{p}]^2}$,  $I_{pp}(t)=\braket{\hat{p}(t)\hat{p}^2\hat{p}(t)}$,  $|\mr{Re}[F_{pp}(t)]|=|\mr{Re}[\braket{\hat{p}(t)\hat{p}\hat{p}(t)\hat{p}}]|,$ and $D_{pp}(t)=\braket{\hat{p}\hat{p}(t)^2\hat{p}}$.
While $D_{pp}(t)$ and $\mr{Re}[F_{pp}(t)]$ behave diffusively and proportional to $t$, $C_{pp}(t)$ and $I_{pp}(t)$ are almost equal and  asymptotically proportional to $t^2$.
We note that the long-time behavior is qualitatively different from that  in Fig.~\ref{figure1}(a) because our model is periodically driven and the momentum is unbounded. We also note that the dynamical localization~\cite{Chirikov88} does not occur within the time scale of our interest.

Figure~\ref{qkr} (b) shows time evolutions of $s_{t},r_{t}$ and $\sqrt{s_{t}r_{t}}$, showing the validity of two conditions (\ref{jo1}) for $t\gtrsim 3$.
Schematically, these conditions are understood from the dynamics of the coarse-grained Wigner function~\cite{TWA-OTOC}, as shown in Fig.~\ref{modoru2}.

Our results demonstrate that Eq.~(\ref{atarasi}) with the conditions~(\ref{jo1}) leads to nontrivial consequences.
Before the Ehrenfest time $t_\mr{E}$ ($t_\mr{E}\simeq 4$ for the parameters in Fig.~\ref{qkr})~\cite{Ehr-OTOC}, $C_{pp}(t)$ grows exponentially by the semiclassical approximation~\cite{Larkin69,Kitaev15S,Cotler17,Rozenbaum17}.
Hence, from the equivalence $C_{pp}(t)\simeq I_{pp}(t)$, irreversibility $I_{pp}(t)$ grows exponentially at short times~\cite{Sch-OTOC}, as demonstrated in Appendix IIIA~\cite{footOTOC1}.
Conversely, for longer times, $I_{pp}(t)\propto t^2$ should follow on physical grounds, as discussed in Appendix IIIB~\cite{footOTOC1}.
Then, $C_{pp}(t)\simeq I_{pp}(t)$  leads  to $C_{pp}(t)\propto t^2$, which has not been calculated from the semiclassical approximation before~\cite{Jac-OTOC}.

We can also analytically discuss the validity of the equivalence of noncommutativity and irreversibility in the semiclassical limit $\hbar_\mr{eff}\ra 0$.
As detailed in Appendix IV~\cite{footOTOC1}, the result depends on the timescale and the perturbation $\hat{A}$.

\paragraph{Conclusion.}
We have found that two distinct probes of quantum chaos, namely noncommutativity  $C_{AB}(t)$ and irreversibility $I_{AB}(t)$, are essentially equivalent to each other for initially localized states (see Eqs.~(\ref{main})-(\ref{jo1})).
We have  verified this for  nonintegrable quantum many-body systems and a quantum kicked rotor.
We have shown that the growth of $C_{AB}(t)$ is dominated by the 3- rather than 4-OTOC for initially localized states.
It is hoped that our finding motivates further experimental study on the OTOC, since nonequilibrium states can be prepared by the quantum quench and $I_{AB}(t)$ can be measured as the expectation value after the time-reversal test~\cite{Garttner17}, which is easier than the measurement of the 4-OTOC.

\begin{acknowledgments}
We are grateful to Naoto Tsuji for insightful discussions at the early stage of our work and helpful comments on the manuscript.
We also thank Kohei Kawabata for carefully reading the manuscript with valuable comments.
We also thank Keiju Murata, Keiji Saito, Naoto Shiraishi, Naoto Kura, and Y\^uto Murashita for fruitful discussions.
This work was supported by
KAKENHI Grant No. JP18H01145 and
a Grant-in-Aid for Scientific Research on Innovative Areas ``Topological Materials Science’’ (KAKENHI Grant No. JP15H05855)
from the Japan Society for the Promotion of Science.
R. H. was supported by the Japan Society for the Promotion of Science through Program for Leading Graduate Schools (ALPS) and JSPS fellowship (JSPS KAKENHI Grant No. JP17J03189).
K. F. was supported by JSPS fellowship (JSPS KAKENHI Grant No. JP16J01683).
\end{acknowledgments}

\bibliography{../../../../refer_them}

\begin{thebibliography}{94}
\bibitem [{Los()}]{Los-OTOC}%
  \BibitemOpen
  \href@noop {} {}\bibinfo {note} {{This is similar to the Loschmidt
  echo~\cite{Peres84S,Jalabert01,Gorin06,Jacquod09}, which measures the
  fidelity between the initial and final states under the time-reversal test.
  In addition, another type of the fidelity, called the displacement echo, is
  also studied in Ref.~\cite{Petitjean07}, where a unitary perturbation is
  added upon reversal.}}\BibitemShut {Stop}%
\bibitem [{Sch()}]{Sch-OTOC}%
  \BibitemOpen
  \href@noop {} {}\bibinfo {note} {{In Ref.~\cite{Schmitt18}, the authors
  discussed an exponential growth of irreversibility via the semiclassical
  approximation of the double commutator in the form of
  $[\hat{X}(t),[\hat{X}(t),\hat{Y}]]$. In contrast, our work relates
  $I_{AB}(t)$ to the more familiar indicator $C_{AB}(t)$, where the squared
  form rules out the vanishment of the chaotic growth upon the
  averaging.}}\BibitemShut {Stop}%
\bibitem [{Cor()}]{Cor-OTOC}%
  \BibitemOpen
  \href@noop {} {}\bibinfo {note} {{We call the correlator as $n$-point if it
  has the form of
  $\braket{\lrl{\prod_{i_1=1}^{k_1}\hat{X}_{i_1}(t_1)}\cdots\lrl{\prod_{i_n=1}^{k_n}\hat{X}_{i_n}(t_n)}}$
  with $t_l\neq t_{l+1}\:(1\leq l<n)$ for an initial state
  $\hat{\rho}=\hat{\rho}(0)$.}}\BibitemShut {Stop}%
\bibitem [{Cha()}]{Change-OTOC}%
  \BibitemOpen
  \href@noop {} {}\bibinfo {note} {{We often consider the case where
  $\braket{\hat{B}}\neq 0$ for $C_{AB}(t)$. If $\braket{\hat{B}}\neq 0$, we
  take $\hat{B}-\braket{\hat{B}}$ as a new $\hat{B}$, which does not change the
  value of $C_{AB}(t)$. Similarly, if the original $\hat{A}$ does not satisfy
  $\braket{\hat{A}}_\beta= 0$, we can shift $\hat{A}$ as
  $\hat{A}\ra\hat{A}-\braket{\hat{A}}_\beta$, which does not change the value
  of $C_{AB}(t)$.}}\BibitemShut {Stop}%
\bibitem [{Per()}]{Pert-OTOC}%
  \BibitemOpen
  \href@noop {} {}\bibinfo {note} {{Here the perturbation is implemented by a
  sudden change of the state, which is different from the usual Loschmidt echo
  protocol~\cite{Gorin06}. In fact, $\hat{\widetilde{\rho}}_{t=0}\neq
  \hat{\rho}$ even for $t=0$.}}\BibitemShut {Stop}%
\bibitem [{Mac()}]{Macro-OTOC}%
  \BibitemOpen
  \href@noop {} {}\bibinfo {note} {{Thermal initial states can be
  microscopically irreversible at the level of density matrices, but they are
  not macroscopically irreversible for small perturbations because the
  expectation value of $\hat{B}^\dag\hat{B}$ does not grow under the
  time-reversal test.}}\BibitemShut {Stop}%
\bibitem [{ETH()}]{ETH-OTOC}%
  \BibitemOpen
  \href@noop {} {}\bibinfo {note} {{The ETH does not hold for integrable
  systems~\cite{Rigol08,Santos10a}, many-body localized
  systems~\cite{Pal10,Nandkishore14,Choi16}, and systems that possess certain
  symmetry sectors~\cite{Hamazaki16G,Shiraishi17,Mori17}.}}\BibitemShut {Stop}%
\bibitem [{TWA()}]{TWA-OTOC}%
  \BibitemOpen
  \href@noop {} {}\bibinfo {note} {{We consider discrete Wigner
  functions~\cite{Kolovsky96}. This process generates unphysical ghost images,
  but they mostly vanish after coarse-graining. We note that while quantum
  interference patterns may vanish by coarse-graining, it does not affect our
  main results.}}\BibitemShut {Stop}%
\bibitem [{Ehr()}]{Ehr-OTOC}%
  \BibitemOpen
  \href@noop {} {}\bibinfo {note} {{Here, the Ehrenfest time is given by
  $t_\mr{E}=\frac{1}{\lambda}\log\frac{S}{\hbar_\mr{eff}}$, where $S$ is some
  characteristic action, $\lambda$ is a constant similar to (but slightly
  larger than~\cite{Rozenbaum17}) the Lyapunov exponent. For the parameters in
  Fig.~4, $S\simeq 30$ and $\lambda\simeq 2$, leading to $t_E\simeq
  3.8$.}}\BibitemShut {Stop}%
\bibitem [{Jac()}]{Jac-OTOC}%
  \BibitemOpen
  \href@noop {} {}\bibinfo {note} {{Several quantities can be calculated from
  the semiclassical approximation even after the Ehrenfest
  time~\cite{Heller93}. The semiclassical theory and the Ehrenfest time are
  also discussed in the study of quantum chaotic
  transport~\cite{Jacquod06}.}}\BibitemShut {Stop}%
\end{thebibliography}
\end{document}


\title{
Supplemental Material for ``Operator Noncommutativity and Irreversibility in Quantum Chaos"}
\date{\today}
\author{Ryusuke Hamazaki}
\affiliation{
Department of Physics, University of Tokyo, 7-3-1 Hongo, Bunkyo-ku, Tokyo 113-0033, Japan
}
\author{Kazuya Fujimoto}
\affiliation{
Department of Physics, University of Tokyo, 7-3-1 Hongo, Bunkyo-ku, Tokyo 113-0033, Japan
}
\author{Masahito Ueda}
\affiliation{
Department of Physics, University of Tokyo, 7-3-1 Hongo, Bunkyo-ku, Tokyo 113-0033, Japan
}
\affiliation{
RIKEN Center for Emergent Matter Science (CEMS), Wako 351-0198, Japan
}

\pacs{05.30.-d, 05.45.Mt}

\maketitle

\section{Proof of some relations about the correlators including Eq.~(1) in the main text}
We start from
\aln{
C_{AB}(t)=I_{AB}(t)+D_{AB}(t)-2\mr{Re}[F_{AB}(t)],
}
where
\aln{
C_{AB}(t)&= \braket{|[\hat{A}(t),\hat{B}]|^2},\nonumber\\
I_{AB}(t)&= \braket{\hat{A}^\dag(t)\hat{B}^\dag\hat{B}\hat{A}(t)},\nonumber\\
D_{AB}(t)&=\braket{\hat{B}^\dag\hat{A}^\dag(t)\hat{A}(t)\hat{B}}, \nonumber\\
F_{AB}(t)&=\braket{\hat{A}^\dag(t)\hat{B}^\dag\hat{A}(t)\hat{B}}.
}
In the following, we omit the subscript $AB$ to simplify the notation.
We first note that, by the Cauchy-Schwarz inequality
\aln{
|F|&=|\Tr[\hrho\hat{A}^\dag(t)\hat{B}^\dag\hat{A}(t)\hat{B}]|\nonumber\\
&=|\Tr[\hrho^{1/2}\hat{A}^\dag(t)\hat{B}^\dag\hat{A}(t)\hat{B}\hrho^{1/2}]|\nonumber\\
&\leq\sqrt{\Tr[\hrho^{1/2}\hat{A}^\dag(t)\hat{B}^\dag\hat{B}\hat{A}(t)\hrho^{1/2}]\Tr[\hrho^{1/2}\hat{B}^\dag\hat{A}(t)^\dag\hat{A}(t)\hat{B}\hrho^{1/2}]}\nonumber\\
&=\sqrt{ID},
}
we can define $\hat{\rho}^{1/2}$ due to the positive definiteness of $\hat{\rho}$~\footnote{Note that  we cannot discuss the difference between $I_{AB}(t)$ and $D_{AB}(t)$ on the right-hand side from the similar inequality for the regularized OTOC for the thermal state~\cite{Maldacena16}.}.

Then
\aln{
(C-I-D)^2&=4\mr{Re}[F]^2\nonumber\\
&\leq 4|F|^2\nonumber\\
&\leq 4ID,
}
and thus
\aln{\label{siki0}
C^2+I^2+D^2-2CD-2ID-2CI\leq 0.
}
Then, we obtain the following inequalities
\aln{
|\sqrt{C}-\sqrt{I}|&\leq \sqrt{D},\label{siki1}\\
|\sqrt{C}-\sqrt{D}|&\leq \sqrt{I},\\
|\sqrt{D}-\sqrt{I}|&\leq \sqrt{C}.
}
Here, we can prove Eq.~(S-\ref{siki1}) by explicitly solving Eq. (S-\ref{siki0}) as a quadratic equation of $D$ under the condition that $C, I, D$ are positive.
Similarly, other inequalities are derived.

From these inequalities, we obtain several results.
For instance, dividing Eq.~(S-\ref{siki1}) by $\sqrt{I}$ leads to
\aln{
\lrv{\sqrt{\frac{C}{I}}-1}\leq\sqrt{\frac{D}{I}},
}
which reduces to Eq.~(1) in the main text.
Similarly, we also obtain the following inequalities:
\aln{
\lrv{\sqrt{\frac{I}{C}}-1}&\leq\sqrt{\frac{D}{C}},\label{sikidaiji1}\\
\lrv{\sqrt{\frac{C}{D}}-1}&\leq\sqrt{\frac{I}{D}},\\
\lrv{\sqrt{\frac{D}{C}}-1}&\leq\sqrt{\frac{I}{C}},\\
\lrv{\sqrt{\frac{D}{I}}-1}&\leq\sqrt{\frac{C}{I}},\\
\lrv{\sqrt{\frac{I}{D}}-1}&\leq\sqrt{\frac{C}{D}}.\label{sikidaiji2}
}
For example, Eq.~(S-\ref{sikidaiji1}) means that noncommutativity $C$ and irreversibility $I$ are equivalent if noncommutativity is much larger than the time-ordered correlation function.
We will use some of these inequalities in Appendix~\ref{semilimit}.


\section{Details for quantum many-body systems}
\subsection{Unchanged energy after the perturbation of $\hat{B}$ on the initial state}
In this subsection, we show that the energies of $\hat{\rho}$ and $\hat{\rho}'=\frac{\hat{B}\hat{\rho}\hat{B}^\dag}{\Tr[{\hat{\rho}\hat{B}^\dag\hat{B}}]}$ are macroscopically equal when $\hat{B}$ can be written as a sum of local operators and $\hat{\rho}$ satisfies the  cluster decomposition property.
We first note $\Tr[\hat{\rho}'\hat{H}]=\frac{\braket{\hat{B}^\dag\hat{H}\hat{B}}}{\braket{\hat{B}^\dag\hat{B}}}$.
Using the cluster decomposition, we can decompose $\braket{\hat{B}^\dag\hat{H}\hat{B}}=\sum_{ijk}\braket{\hat{b}_i^\dag\hat{h}_j\hat{b}_k}$ into
\aln{\label{mobu}
\sum_{(i,j,k)\in \mc{A}_1}\braket{\hat{b}_i^\dag\hat{h}_j\hat{b}_k}+\sum_{(i,j,k)\in \mc{A}_2}\braket{\hat{b}_i^\dag\hat{b}_k}\braket{\hat{h}_j}.
}
Here, $\mc{A}_1$ is a set of trios $(i,j,k)$ where $i,j$ and $k$ are close to each other (i.e., their distance is independent of the size of the system), and $\mc{A}_2$ is a set of trios where $i$ and $k$ are close to each other but neither of them is close to $j$.
Note that the contributions from other trios vanish due to $\braket{\hat{b}_i}=0$, which results from $\braket{\hat{B}}=0$ as described in the main text.
The first term on the right-hand side of Eq. (S-\ref{mobu}) is proportional to $N$ while the second one is of the order of $N^2$. Thus, the leading term of Eq.~(S-\ref{mobu}) is estimated to be
\aln{
\braket{\hat{H}}\sum_{(i,k)\in\mc{A}_0}\braket{\hat{b}_i^\dag\hat{b}_k}.
}
Here, $\mc{A}_0$ is a set of pairs $(i,k)$ where $i$ and $k$ are close to each other and we can safely replace $\sum_{(i,j,k)\in \mc{A}_2}$ with $\sum_j\sum_{(i,k)\in\mc{A}_0}$ without changing the leading contribution.
Similarly, the denominator is approximated by $\sum_{(i,k)\in\mc{A}_0}\braket{\hat{b}_i^\dag\hat{b}_k}$ up to the leading order, giving
\aln{
\Tr[\hat{\rho}'\hat{H}]\simeq \braket{\hat{H}}=\Tr[\hat{\rho}\hat{H}]
}
in the thermodynamic limit.

\subsection{Behavior of $r_t$ for many-body systems}
To derive the explicit formula for $r_t$, we use several physically reasonable assumptions, in addition to the ones mentioned in the main text (e.g., the ETH and the translational invariance).
For initial states, 
we assume that the it has most of its support on the microcanonical energy shell, i.e., $\rho_{\gamma\alpha}$ is negligibly small if $\ket{E_\alpha}$ or $\ket{E_\gamma}$ lies outside the microcanonical shell.
This is justified for the typical quench protocol~\cite{Rigol08}.
For the operator $\hat{A}$, we assume that it can be written as a local operator or a sum of local operators.
For the Hamiltonian $\hat{H}$, we assume that energy eigenvalues and their difference are not degenerate~\cite{Reimann08}.

The long-time behavior of $r_t^{-1}$ can explicitly be calculated as
\aln{
r_t^{-1}&=\frac{\Tr[\hat{\widetilde{\rho}}_t\hat{B}^\dag\hat{B}]}{\Tr[\hat{\rho}\hat{B}^\dag\hat{B}]}\NON
&=\frac{\braket{\hat{A}(t)^\dag\hat{B}^\dag\hat{B}\hat{A}(t)}}{\braket{\hat{B}^\dag\hat{B}}\braket{\hat{A}^\dag(t)\hat{A}(t)}}\NON
&\simeq \frac{\braket{\hat{A}(t)^\dag\hat{B}^\dag\hat{B}\hat{A}(t)}}{\braket{\hat{B}^\dag\hat{B}}\braket{\hat{A}^\dag\hat{A}}_\beta},
}
where we have replaced $\braket{\hat{A}^\dag(t)\hat{A}(t)}$ with $\braket{\hat{A}^\dag\hat{A}}_\beta$ in the long-time dynamics, since $\braket{\hat{A}^\dag(t)\hat{A}(t)}\simeq \braket{\hat{A}^\dag\hat{A}}_\beta$ for most of the times with negligible fluctuations due to the eigenstate thermalization hypothesis (ETH)~\cite{Rigol08,Reimann08}.
Similarly, we also assume that the long-time behavior of the numerator has negligibly small temporal fluctuations around the averaged value.
Then, using the assumption about the non-degeneracy of energy eigenvalues and gaps~\cite{Reimann08}, we obtain
\aln{
r_t^{-1}\simeq  \frac{\sum_{\alpha\neq\gamma}\rho_{\gamma\alpha}A_{\alpha\alpha}^*A_{\gamma\gamma}(\hat{B}^\dag\hat{B})_{\alpha\gamma}+\sum_{\alpha\gamma}\rho_{\alpha\alpha}|A_{\alpha\gamma}|^2(\hat{B}^\dag\hat{B})_{\gamma\gamma}}{\braket{\hat{B}^\dag\hat{B}}\braket{\hat{A}^\dag\hat{A}}_\beta},
}
where we define the matrix elements with respect to energy eigenstates $\ket{E_\alpha}$ as  $A_{\alpha\gamma}=\braket{E_\alpha|\hat{A}|E_\gamma}$, etc.

Next, from the assumption for $\hat{A}$,
the matrix elements of $A_{\alpha\gamma}$ are shown to be suppressed exponentially for large $|E_\alpha-E_\gamma|$ as $|A_{\alpha\gamma}|\propto e^{-c|E_\alpha-E_\gamma|}$ with some constant $c$~\cite{Abanin15}.
Thus, we can approximately replace the sum over $\alpha$ and $\gamma$ of the entire Hilbert space with the sum over $\alpha$ and $\gamma$ within the microcanonical energy shell (and vice versa).
It follows from the ETH that
\aln{
A_{\alpha\alpha}\simeq A_{\gamma\gamma}&\simeq \braket{\hat{A}}_\beta,\\
(\hat{B}^\dag\hat{B})_{\gamma\gamma}&\simeq \braket{\hat{B}^\dag\hat{B}}_\beta,
}
where $\beta$ is the inverse temperature that corresponds to the energy of the initial state.
Then, denoting the sum over eigenstates in the microcanonical ensemble as $\sum'$, we obtain
\aln{
r_t^{-1}&\simeq  \frac{\sum_{\alpha\neq\gamma}'\rho_{\gamma\alpha}A_{\alpha\alpha}^*A_{\gamma\gamma}(\hat{B}^\dag\hat{B})_{\alpha\gamma}+\sum_{\alpha}\sum_\gamma'\rho_{\alpha\alpha}|A_{\alpha\gamma}|^2(\hat{B}^\dag\hat{B})_{\gamma\gamma}}{\braket{\hat{B}^\dag\hat{B}}\braket{\hat{A}^\dag\hat{A}}_\beta} \NON
&\simeq  \frac{|\braket{\hat{A}}_\beta|^2\sum_{\alpha\neq\gamma}'\rho_{\gamma\alpha}(\hat{B}^\dag\hat{B})_{\alpha\gamma}+ \braket{\hat{B}^\dag\hat{B}}_\beta\sum_{\alpha}\sum_\gamma'\rho_{\alpha\alpha}|A_{\alpha\gamma}|^2}{\braket{\hat{B}^\dag\hat{B}}\braket{\hat{A}^\dag\hat{A}}_\beta} \NON
&\simeq  \frac{|\braket{\hat{A}}_\beta|^2\sum_{\alpha\neq\gamma}\rho_{\gamma\alpha}(\hat{B}^\dag\hat{B})_{\alpha\gamma}+ \braket{\hat{B}^\dag\hat{B}}_\beta\sum_{\alpha}\sum_\gamma\rho_{\alpha\alpha}|A_{\alpha\gamma}|^2}{\braket{\hat{B}^\dag\hat{B}}\braket{\hat{A}^\dag\hat{A}}_\beta} \NON
&\simeq  \frac{|\braket{\hat{A}}_\beta|^2\sum_{\alpha\neq\gamma}\rho_{\gamma\alpha}(\hat{B}^\dag\hat{B})_{\alpha\gamma}+ \braket{\hat{B}^\dag\hat{B}}_\beta\sum_{\alpha}\rho_{\alpha\alpha}(\hat{A}^\dag\hat{A})_{\alpha\alpha}}{\braket{\hat{B}^\dag\hat{B}}\braket{\hat{A}^\dag\hat{A}}_\beta}\NON
&\simeq\frac{|\braket{\hat{A}}_\beta|^2(\braket{\hat{B}^\dag\hat{B}}-\braket{\hat{B}^\dag\hat{B}}_{\beta})+ \braket{\hat{B}^\dag\hat{B}}_\beta\braket{\hat{A}^\dag\hat{A}}_\beta}{\braket{\hat{B}^\dag\hat{B}}\braket{\hat{A}^\dag\hat{A}}_\beta}\NON
&= \frac{|\braket{\hat{A}}_\beta|^2}{\braket{\hat{A}^\dag\hat{A}}_\beta}+\lrs{1-\frac{|\braket{\hat{A}}_\beta|^2}{\braket{\hat{A}^\dag\hat{A}}_\beta}}\frac{\braket{\hat{B}^\dag\hat{B}}_\beta}{\braket{\hat{B}^\dag\hat{B}}},
}
where we have used
\aln{
\sum_{\alpha\neq\gamma}\rho_{\gamma\alpha}(\hat{B}^\dag\hat{B})_{\alpha\gamma}&=\sum_{\alpha\gamma}\rho_{\gamma\alpha}(\hat{B}^\dag\hat{B})_{\alpha\gamma}-\sum_\alpha\rho_{\alpha\alpha}(\hat{B}^\dag\hat{B})_{\alpha\alpha}\NON
&\simeq\braket{\hat{B}^\dag\hat{B}}-\braket{\hat{B}^\dag\hat{B}}_{\beta}
}
with the help of the ETH.

For initially localize states, $\Tr[\hat{\rho}\hat{B}^\dag\hat{B}]=\braket{\hat{B}^\dag\hat{B}}\ll\braket{\hat{B}^\dag(t)\hat{B}(t)}\simeq \braket{\hat{B}^\dag\hat{B}}_{\beta}$ because of their definition and the ETH,
and thus $r_t\ll 1$ in the thermodynamic limit if $1-\frac{|\braket{\hat{A}}_\beta|^2}{\braket{\hat{A}^\dag\hat{A}}_\beta}$ is $\mr{O}(N^0)$.
This condition trivially holds true if we assume that $\braket{\hat{A}}_\beta=0$.
If this is not the case, we can define $\hat{A}-\braket{\hat{A}}_\beta$ as new $\hat{A}$ without changing the value of $C_{AB}(t)$.
Moreover,  the condition holds true without this procedure if $\hat{A}$ is a  typical local operator such as the Pauli operators $\hat{\sigma}_i^{x,y,z}$, where $\frac{|\braket{\hat{A}}_\beta|^2}{\braket{\hat{A}^\dag\hat{A}}_\beta}=|\braket{\hat{\sigma}_i^{x,y,z}}_\beta|^2$ is smaller than 1 in typical situations.

\section{Details of numerical simulations of the quantum kicked rotor}
\subsection{Short-time behavior and semiclassical representation}
In this section, we consider the short-time dynamics of the quantum kicked rotor (Eq.~(6) in the main text) before the Ehrenfest time $t_\mr{E}$ and its semiclassical representation.
We consider two localized initial states in momentum space (i.e., $\hat{B}=\hat{p}$).
The first is a wave-packet state
\aln{
\hat{\rho}_\mr{w}:=\ket{\psi_\mr{w}}\bra{\psi_\mr{w}},\:\: \ket{\psi_\mr{w}}:=\frac{1}{Z_\mr{w}}\sum_m e^{-\frac{\hbar_\mr{eff} m^2}{2\sigma^2}}\ket{p_m}\:\:\:\lrs{Z_\mr{w}:=\sqrt{\sum_m e^{-\frac{\hbar_\mr{eff}m^2}{\sigma^2}}}},
}
which is discussed in the main text.
The second is the canonical distribution for a free Hamiltonian
$\hat{H}_0:=\frac{\hat{p}^2}{2},$
\aln{
\hat{\rho}_T:=\frac{1}{Z_T}\sum_m e^{-\frac{\hbar_\mr{eff}^2m^2}{2T}}\ket{p_m}\bra{p_m}\:\:\:\lrs{Z_T:=\sum_m e^{-\frac{\hbar_\mr{eff}^2m^2}{2T}}}.
}
This initial canonical distribution is localized with respect to $\hat{p}$ (but not $\hat{x}$) when we consider the Floquet time evolution $\hat{F}=e^{-\frac{i\hat{p}^2}{2\hbar_\mr{eff}}}e^{-\frac{iK\cos\hat{x}}{\hbar_\mr{eff}}}$ for $\hat{H}(t)$ (see Eq. (6) in the main text).
Note that the state is not stationary ($[\hat{F},\hat{\rho}_T]\neq 0$) due to periodic kicks.
Thus, $I_{AB}(t)=\braket{\hat{A}^\dag(t)\hat{B}^\dag\hat{B}\hat{A}(t)}$ becomes a 3-OTOC for these initial states.

For reference, we also apply a semiclassical approximation to each correlator.
We consider the average of a classical function $\mc{S}(x,p,t)$ over the Wigner distribution $W$ of the initial states,
\aln{
\av{\mc{S}_t}:=\int dxdp W(x,p)\mc{S}(x,p,t).
}
As shown below, every correlator is approximated before $t_\mr{E}$  by $\av{\mc{S}_t}$ for an appropriate
$\mc{S}(x,p,t)$.
The Wigner distributions of our initial states, $\hat{\rho}_\mr{w}$ and $\hat{\rho}_T$, are approximated in Gaussian forms as
\aln{
W_\mr{w}(x,p)=\frac{1}{\pi\hbar_\mr{eff}}e^{-\frac{p^2}{\hbar_\mr{eff}\sigma^2}-\frac{\sigma^2x^2}{\hbar_\mr{eff}}}
}
 and
\aln{
W_{T}(x,p)=\frac{1}{\sqrt{(2\pi)^3T}}e^{-\frac{p^2}{2T}},
}
 respectively~\footnote{Here we ignore the discreteness of $p$, which is justified for small $\hbar_\mr{eff}$. Because of this, the periodicity about $x$ is lost. In particular, unphysical ghost images in Ref.~\cite{Kolovsky96} are lost, which are expected not to change the results for small $\hbar_\mr{eff}$.}.

\begin{figure}
\begin{center}
\includegraphics[width=\linewidth]{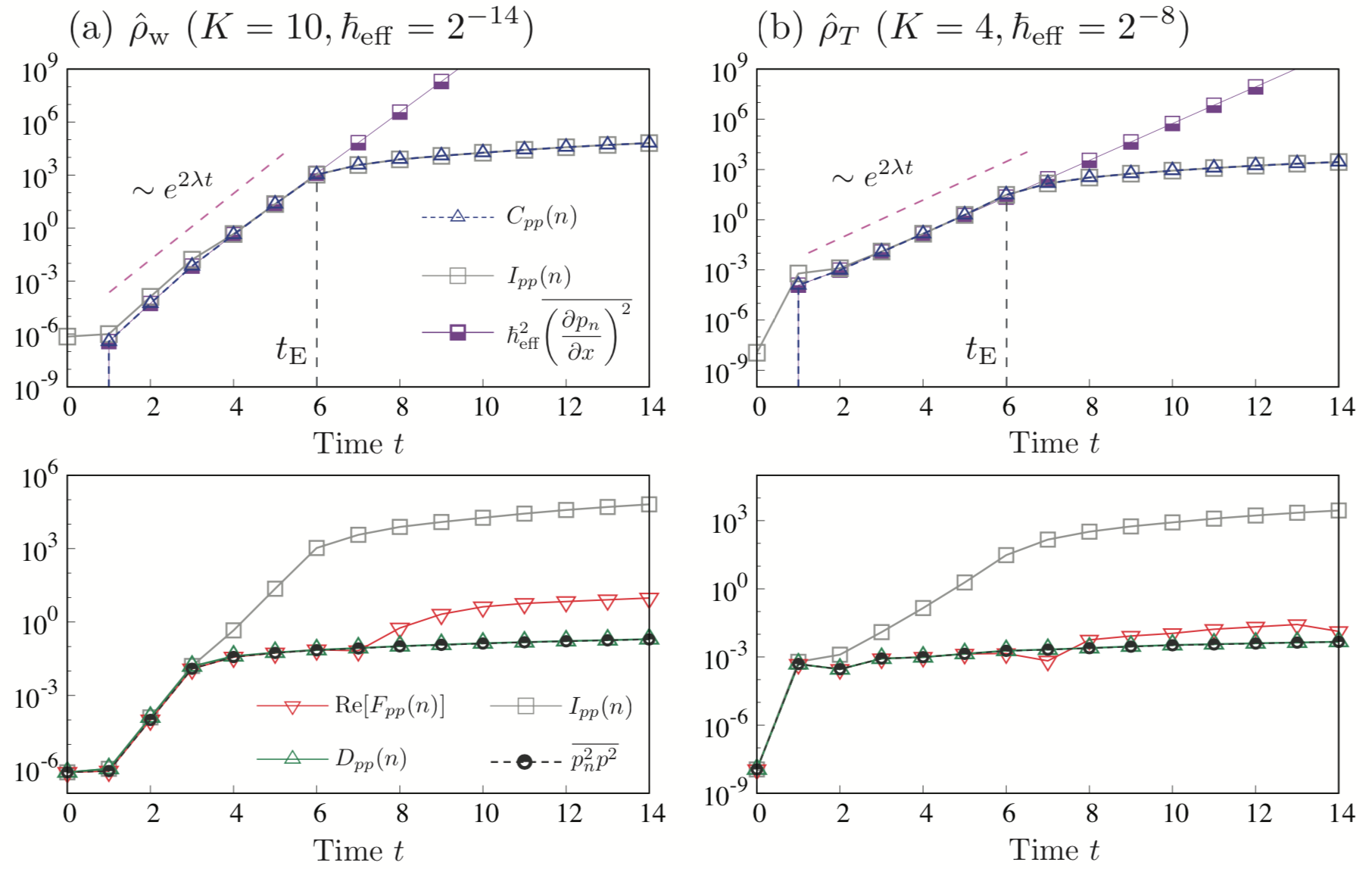}
\caption{Short-time dynamics of $C_{pp}(t)=-\braket{[\hat{p}(t),\hat{p}]^2}$, $D_{pp}(t)=\braket{\hat{p}\hat{p}(t)^2\hat{p}}$, $\mr{Re}[F_{pp}(t)]=\mr{Re}[\braket{\hat{p}(t)\hat{p}\hat{p}(t)\hat{p}}]$, $I_{pp}(t)=\braket{\hat{p}(t)\hat{p}^2\hat{p}(t)}$, $\av{p_t^2p^2}$, and $\hbar_\mr{eff}^2\av{\lrs{\fracpd{p_t}{x}}^2}$
 for initial states (a) $\hat{\rho}_\mr{w}$ and  (b) $\hat{\rho}_T$.
(upper panels) For both initial states and up to the Ehrenfest time $t\lesssim t_E\sim6$, $C_{pp}(t)$ and $I_{pp}(t)$ agree excellently and grow exponentially, and they are well approximated by $\hbar_\mr{eff}^2\av{\lrs{\fracpd{p_t}{x}}^2}$.
(bottom panels) Both  $D_{pp}(t)$ and $\mr{Re}[F_{pp}(t)]$ are well described by the  classical average $\av{p_t^2p^2}$  for $t\lesssim t_E$, unlike $I_{pp}(t)$.
}
\label{yonten}
\end{center}
\end{figure}

As shown in Fig.~\ref{yonten}, we first consider the short-time behaviors of $C_{pp}(t)=-\braket{[\hat{p}(t),\hat{p}]^2}$,  $I_{pp}(t)=\braket{\hat{p}(t)\hat{p}^2\hat{p}(t)}$ ($\hat{p}(t):=(\hat{F}^\dag)^n\hat{p}\hat{F}^n$),  $\mr{Re}[F_{pp}(t)]=\mr{Re}[\braket{\hat{p}(t)\hat{p}\hat{p}(t)\hat{p}}],$  $D_{pp}(t)=\braket{\hat{p}\hat{p}(t)^2\hat{p}}$, the classical average neglecting the noncommutativity $\av{p_t^2p^2}$, and the initial sensitivity $\hbar_\mr{eff}^2\av{\lrs{\fracpd{p_t}{x}}^2}$. The left and right figures correspond to $\hat{\rho}_\mr{w}$ and $\hat{\rho}_T$, respectively.
For $t\lesssim t_E\simeq 6$, $D_{pp}(t)$ and $\mr{Re}[F_{pp}(t)]$ are well described by $\av{p_t^2p^2}$, whereas the 3-OTOC $I_{pp}(t)$ grows exponentially.
The exponential growth of $I_{pp}(t)$ represents the initial sensitivity of classical chaos because it is close to $C_{pp}(t)$ (i.e., Eq.~(4) with the conditions (5) in the main text holds true), which reduces to $\hbar_\mr{eff}^2\av{\lrs{\fracpd{p_t}{x}}^2}$~\cite{Larkin69,Kitaev15S,Rozenbaum17} in the semiclassical limit.

\subsection{Origin of the anomalous quadratic scaling in the long-time behavior}

\begin{figure}[t]
\begin{center}
\includegraphics[width=\linewidth]{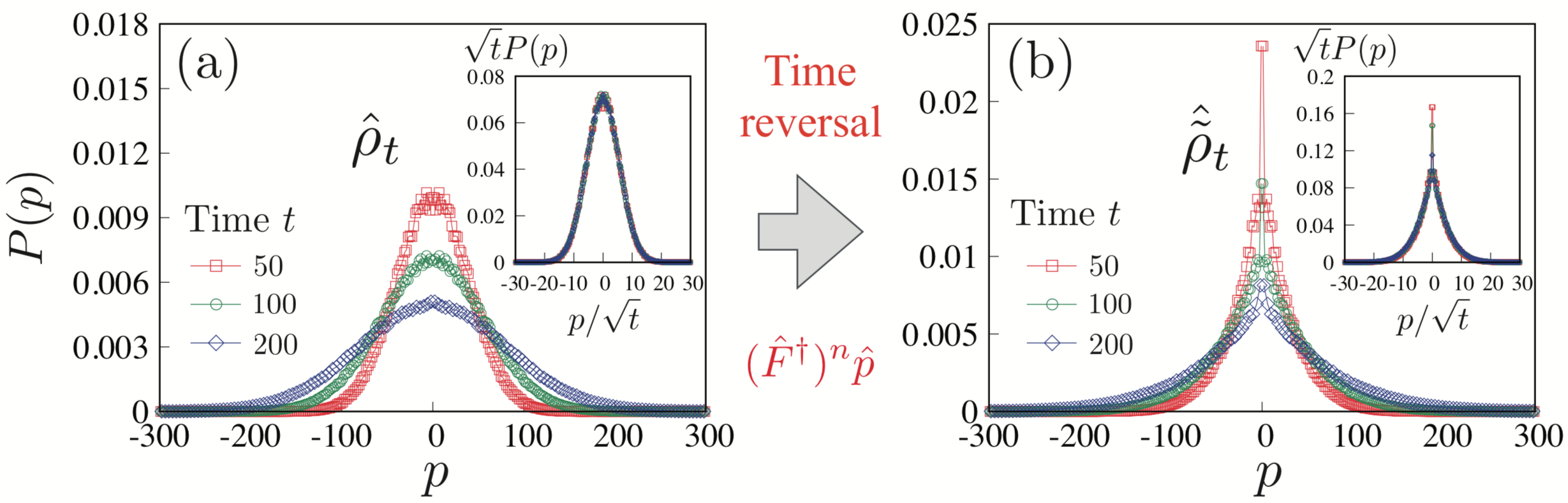}
\caption{
Momentum distributions $P(p)$ of (a) $\hat{\rho}_{t}$ and (b) $\hat{\widetilde{\rho}}_{t}$ for $t=50$, 100 and 200.
For $\hat{\rho}_{t}$, $P(p)$ is close to a Gaussian form and spreads in time.
For $\hat{\widetilde{\rho}}_{t}$, $P(p)$ is not Gaussian, but spreads as $t$ increases.
(insets) Dynamical scaling of $P(p)$ for $\hat{\rho}_{t}$ and $\hat{\widetilde{\rho}}_{t}$.
All the curves collapse to a single curve after rescaling according to $P(p,t)=f(p/\sqrt{t})/\sqrt{t}$ for both $\hat{{\rho}}_{t}$ and $\hat{\widetilde{\rho}}_{t}$.
For both figures, we take the initial state $\hat{\rho}_\mr{w}$ of the wave packet with $\hbar_\mr{eff}=2^{-6},\sigma=4$, and $K=10$.
}
\label{longtime}
\end{center}
\end{figure}

As we have seen in Fig. 4 (a) in the main text, $D_{pp}(t)$ grows diffusively in the long-time regime as $\propto t$ (the dynamical localization~\cite{Chirikov88} does not occur within the time scale of our interest).
Indeed, from Eq.~(2) in the main text, $D_{pp}(t)$ is the product of $\Tr[{\hat{\rho}\hat{p}^2}]$ (where we use the wave-packet state $\hat{\rho}=\hat{\rho}_\mr{w}$) and $\Tr[{{\hat{\rho}'_{t}}\hat{p}^2}]$.
The former does not depend on time and the latter behaves diffusively~\cite{Chirikov88}, so that this time-ordered correlator grows as $\propto t$.

On the other hand, $I_{pp}(t)$ is proportional to $t^2$, which is different from the classical diffusive behavior.
Thanks to Eqs.~(4) and (5) in the main text, $C_{pp}(t)$ also follows a $t^2$ power law~\cite{Rozenbaum17}.
The anomalous quadratic scaling for $I_{pp}(t)$ originates from the fact that the momentum distribution of $\hat{\widetilde{{\rho}}}_{t}$ spreads as much as that of $\hat{{\rho}}_{t}$.
As we have seen in Eq.~(2) in the main text, $I_{pp}(t)$ is the product of
$\Tr[\hat{\widetilde{{\rho}}}_{t}\hat{p}^2]$ and $\Tr[{{\hat{\rho}_{t}}\hat{p}^2}]$.
Figures~\ref{longtime} (a) and (b) plot the coarse-grained momentum distribution
\aln{
P(p):=\frac{1}{\Delta p}\sum_{p_m\in[p-\Delta p/2,p+\Delta p/2)}\braket{p_m|\hat{\rho}|p_m}}
 for $\hat{\rho}_{t}$ and $\hat{\widetilde{\rho}}_{t}$,
respectively.
Figure~\ref{longtime} (a) shows a diffusive, Gaussian profile in quantum chaos~\cite{Altland93}.
After the  time-reversal test, $\hat{\widetilde{\rho}}_{t}$ will remain extended in momentum space, especially for large $t$.
In this time evolution, $P(p)$ obeys
a dynamical scaling relation
\aln{\label{tekitou}
P(p,t)=\frac{1}{\sqrt{t}}f(p/\sqrt{t})
}
 for $\hat{{\rho}}_{t}$ and $\hat{\widetilde{\rho}}_{t}$, as shown in the insets of Fig.~\ref{longtime}(a) and (b)~\footnote{Precisely speaking, we have found a peak at $p=0$, which is not scaled diffusively. However, this peak does not affect the main discussion.}.
Note that $P(p)$ for $\hat{\widetilde{\rho}}_{t}$ obeys the above-mentioned diffusive scaling, even though it is not Gaussian.
Such a delocalization, which obeys the scaling in Eq. (S-\ref{tekitou}), leads to
\aln{
\Tr[\hat{\widetilde{{\rho}}}_{t}\hat{p}^2]\simeq \int dpp^2P(p,t)\propto t.
}
Thus, Eq.~(2) in the main text and the above-mentioned diffusive behavior of $\braket{\hat{p}(t)^2}$ give $\braket{\hat{p}(t)\hat{p}^2\hat{p}(t)}\propto t^2$.
This clearly shows that the 3-OTOC $I_{pp}(t)$ gives the measure of irreversibility that explains the anomalous power-law growth of $C_{pp}(t)$ in the long-time regime.

\subsection{Unitary perturbations}
\begin{figure}
\begin{center}
\includegraphics[width=\linewidth]{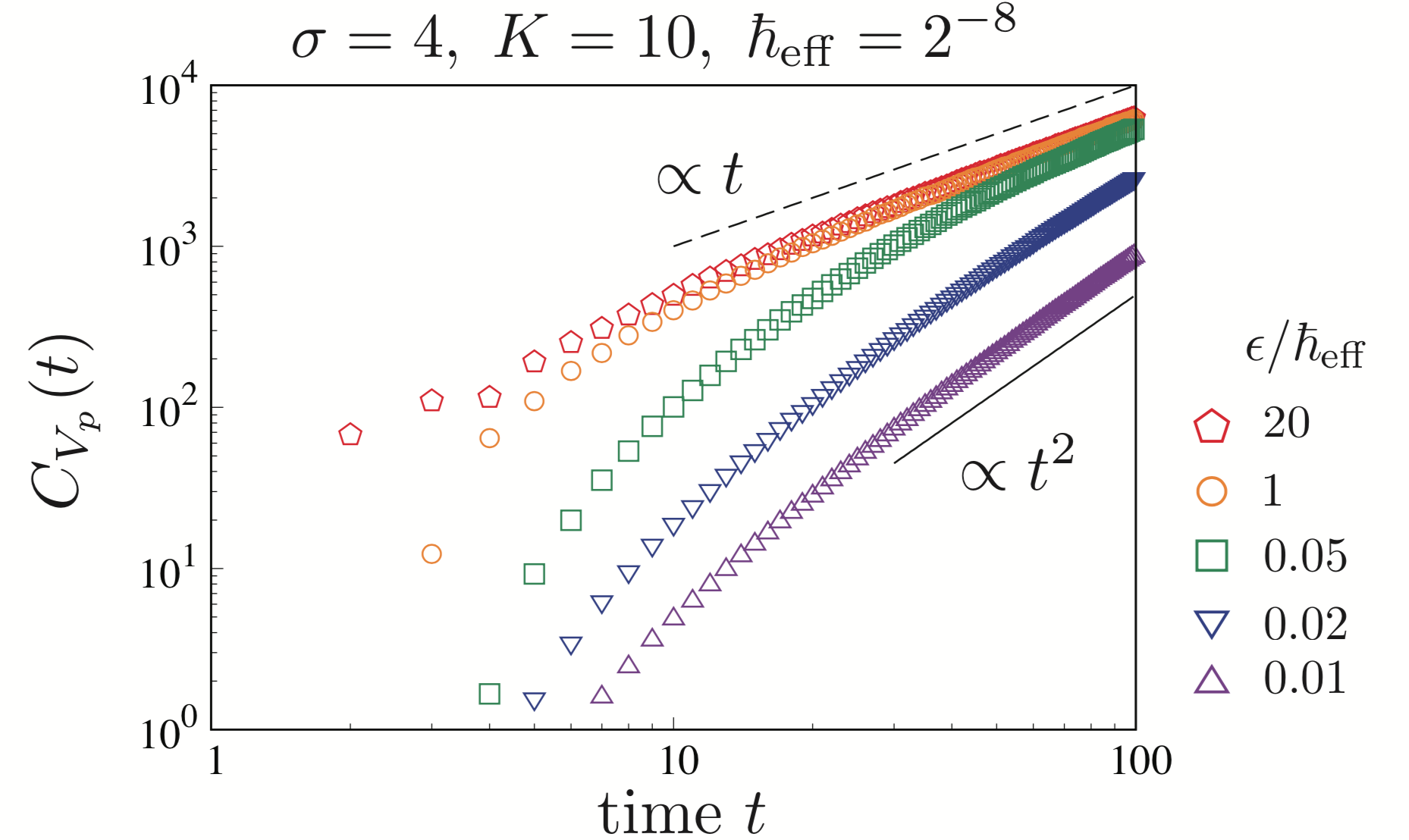}
\caption{
Long-time dynamics of the squared commutator for different strengths of  perturbation: $\epsilon/\hbar=0.01$, 0.02, 0.05, 1, and 20.
For small perturbations, the growth depends on $\epsilon/\hbar$ and exhibits  quadratic scaling.
For strong perturbations, the growth is independent on $\epsilon/\hbar$ and exhibits linear scaling.
}
\label{figure4}
\end{center}
\end{figure}

Here, we consider unitary perturbations, which can often be implemented  experimentally~\cite{Garttner17}.
We take $\hat{A}=\hat{V}=e^{\frac{i\hat{p}\epsilon}{\hbar_\mr{eff}}}$, which translates the state by $\epsilon$ in the $x$ direction, and $\hat{B}=\hat{p}$~\cite{Adachi88,Yamada12}.
Similarly to the case of $\hat{A}=\hat{p}$, the short-time dynamics of
\aln{
C_{Vp}(t)=-\braket{|[e^{\frac{i\hat{p}(t)\epsilon}{\hbar_\mr{eff}}},\hat{p}]|^2}
}
 exhibits an exponential growth that corresponds to $-\hbar_\mr{eff}^2\av{\lrv{\fracpd{}{x}e^{\frac{i{p}_t\epsilon}{\hbar_\mr{eff}}}}^2}=\epsilon^2\av{\lrs{\fracpd{p_t}{x}}^2}$ before $t_\mr{E}$ (data not shown).
On the other hand, as shown in Fig.~\ref{figure4}, $C_{Vp}(t)$ for large $t$ grows as $\propto t^2$ and $\propto t$ for small and large perturbations $\epsilon$, respectively.

The perturbation-dependent behavior can be understood, by using Eq.~(4) with the conditions (5) in the main text, from the behavior of the following 3-OTOC:
\aln{
I_{Vp}(t)=\braket{e^{\frac{-i\hat{p}(t)\epsilon}{\hbar_\mr{eff}}}\hat{p}^2e^{\frac{i\hat{p}(t)\epsilon}{\hbar_\mr{eff}}}}=\braket{\tilde{\psi}_t|\hat{p}^2|\tilde{\psi}_t}\:(\ket{\tilde{\psi}_t}=e^{\frac{i\hat{p}(t)\epsilon}{\hbar_\mr{eff}}}\ket{\psi}).
}
When the perturbation is so small that
\aln{
\frac{\epsilon^2\braket{\hat{p}(t)^2}}{\hbar_\mr{eff}^2}\simeq \frac{\epsilon^2t}{\hbar_\mr{eff}^2}\ll 1\text{ for a given $t$,}
}
we have
\aln{
\ket{\tilde{\psi}_t}\simeq \lrs{1+\frac{i\hat{p}(t)\epsilon}{\hbar_\mr{eff}}}\ket{\psi}
}
 and the dynamics is almost reversible in terms of fidelity ($\braket{\psi|\tilde{\psi}_t}\simeq 1$).
However, $I_{Vp}(t)$ can be approximated as
\aln{
\braket{\tilde{\psi}_t|\hat{p}^2|\tilde{\psi}_t}\simeq \frac{\epsilon^2}{\hbar_\mr{eff}^2}\braket{\hat{p}(t)\hat{p}^2\hat{p}(t)},
}
 which grows in proportion to $t^2$ as can be seen from the results in the previous section~\footnote{The zeroth-order term in $\epsilon$ is time-independent. The first-order terms can be evaluated as $|\frac{i\epsilon}{\hbar_\mr{eff}}(\braket{\hat{p}(t)\hat{p}^2}+\braket{\hat{p}^2\hat{p}(t)})|\leq \frac{2\epsilon\sqrt{\braket{\hat{p}(t)^2}\braket{\hat{p}^4}}}{\hbar_\mr{eff}}\simeq \frac{2\epsilon\sqrt{t\braket{\hat{p}^4}}}{\hbar_\mr{eff}}$, which is small by our assumption}.
In this case, $I_{Vp}(t)$ becomes sufficiently large, providing a measure of irreversibility which is more sensitive than fidelity~\footnote{In Ref.~\cite{Schmitt18}, the authors expand $\braket{\hat{V}^\dag(t)\hat{X}\hat{V}(t)}$ ($\hat{V}=e^{-i\hat{H}\epsilon}$) up to the second order in $\epsilon$, especially before $t_\mr{E}$. Although the expansion series have the same form as ours, we argue that the expansion radius is determined by the expansion for the state, not for the correlator as they discuss.}.
On the other hand, for large perturbation $\frac{\epsilon^2t}{\hbar_\mr{eff}^2}\simeq 1$, the completely irreversible (diffusive) delocalization of $\ket{\tilde{\psi}_t}$ occurs, leading to
\aln{
\braket{e^{\frac{-i\hat{p}(t)\epsilon}{\hbar_\mr{eff}}}\hat{p}^2e^{\frac{i\hat{p}(t)\epsilon}{\hbar_\mr{eff}}}}\propto t.
}
Note that we find a crossover into this regime even for small $\epsilon$ if we wait for a long time (i.e., large $t$).
For both cases, $C_{Vp}(t)\simeq I_{Vp}(t)$ (data not shown) holds true, which leads to results in Fig.~\ref{figure4}.

\section{Semiclassical limits for the quantum kicked rotor}\label{semilimit}
\begin{figure}
\begin{center}
\includegraphics[width=\linewidth]{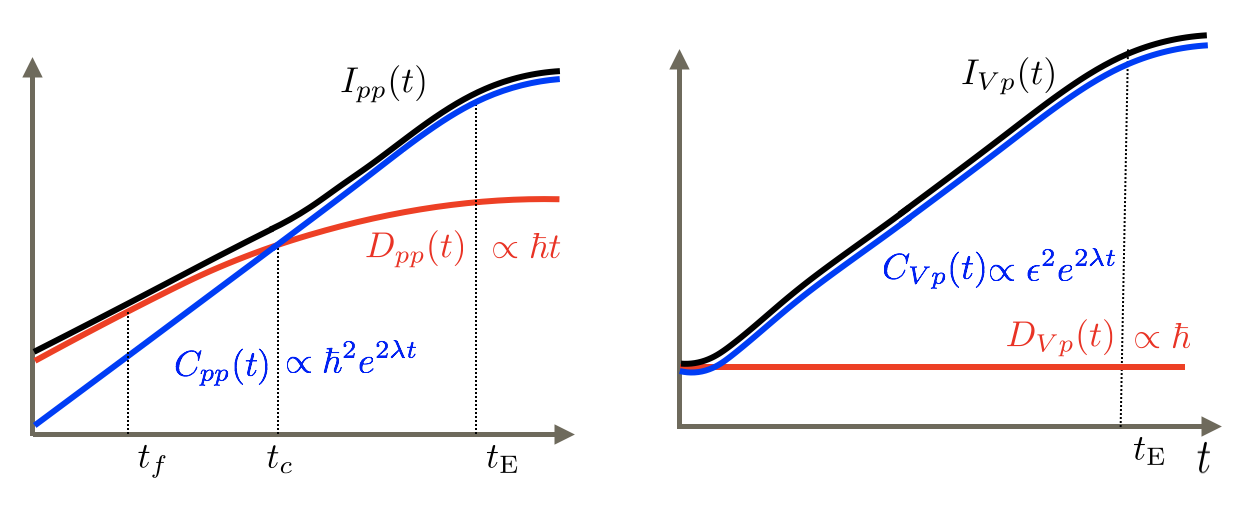}
\caption{
Schematic illustrations for time evolutions of $I_{AB}(t)$, $C_{AB}(t)$ and $D_{AB}(t)$ for sufficiently small $\hbar_\mr{eff}$ (semi-log plot). The left figure shows the case of $\hat{A}=\hat{B}=\hat{p}$. For a fixed time, the approximated equality $I_{pp}(t)\simeq C_{pp}(t)$ is not valid. After some crossover time $t_c$, which grows with $1/\hbar_\mr{eff}$  more slowly than $t_E$, the equivalence $I_{pp}(t)\simeq C_{pp}(t)$ eventually holds. The right figure shows the case of $\hat{A}=\hat{V}=e^{\frac{i\epsilon\hat{p}}{\hbar_\mr{eff}} }$ and $\hat{B}=\hat{p}$. In this case, $I_{Vp}(t)\simeq C_{Vp}(t)$ holds for $\hbar_\mr{eff}\ra 0$ at any timescale.
}
\label{figure7}
\end{center}
\end{figure}

In this section, we discuss the validity of the equivalence between irreversibility $I_{AB}(t)$ and noncommutativity $C_{AB}(t)$ in the semiclassical limit $\hbar_\mr{eff}\ra 0$ for our quantum kicked rotor.
As shown in the following, the validity depends on the timescale  and the types of perturbations, as summarized in Fig.~\ref{figure7}.
In the following, we utilize the inequalities (S-\ref{sikidaiji1}) and (S-\ref{sikidaiji2}), from which we can conclude that
\aln{
I_{AB}(t)&\simeq D_{AB}(t)\:\:\:\text{ when } C_{AB}(t)\ll D_{AB}(t);\label{ahoda1}\\
I_{AB}(t)&\simeq C_{AB}(t)\:\:\:\text{ when } D_{AB}(t)\ll C_{AB}(t).\label{ahoda2}
}

We focus on the localized wave-packet initial state  $\hrho_\mr{w}$ in the main text, where $\braket{\hat{p}^2}\propto\hbar_\mr{eff}$.
We first consider the case of the Hermitian perturbation $\hat{A}=\hat{p}$ and $\hat{B}=\hat{p}$.
In this case, $D_{pp}(t)$ is expected to behave diffusively~\cite{Altland93} as $D_{pp} (t)\simeq a\hbar_\mr{eff} t$, where $a$ is some constant. On the other hand, the semiclassical approximation~\cite{Rozenbaum17} leads to $C_{pp} (t)\simeq b\hbar_\mr{eff}^2 e^{2\lambda t}$, where $b$ is another constant.

Thus, for a fixed time (i.e., an $\hbar_\mr{eff}$-independent time) $t=t_f$ and small $\hbar_\mr{eff}$ limit, $ I_{pp} (t_f )\simeq a\hbar_\mr{eff} t_f$ (since $a\hbar_\mr{eff} t_f \gg b\hbar_\mr{eff}^2 e^{2\lambda t_f }$ and Eq. (S-\ref{ahoda1})), which means the equivalence $I_{AB} (t)\simeq C_{AB} (t)$ is not valid at this timescale and that our initial state is reversible in the semiclassical limit.

For the irreversibility to occur, we require a longer time in this case. To see this, we next consider the Ehrenfest time $t_E\sim \frac{1}{\lambda} \log \frac{1}{\hbar_\mr{eff}}$ ($\lambda$ is a constant similar to the Lyapunov exponent), which slowly diverges for $\hbar_\mr{eff}\ra 0$. In this timescale, we obtain $I_{pp} (t)\simeq C_{pp} (t)$ because of $a\hbar_\mr{eff} t_E  \ll b\hbar_\mr{eff}^2 e^{2λt_E}$ and Eq. (S-\ref{ahoda2}), which means that the equivalence is valid. In fact, there exists a crossover time $t_c$  ($t_f\ll t_c \ll t_E$) that satisfies $D_{pp} (t_c )\simeq C_{pp} (t_c )$. For $t_c\ll t$, $I_{pp} (t)\simeq C_{pp} (t)$ holds true. See Figure~\ref{figure7} (left).

We next discuss the case where the perturbation is unitary $\hat{A}=\hat{V}=e^{\frac{i\epsilon\hat{p}}{\hbar_\mr{eff}} }$. We assume that $\epsilon$ is fixed and $\hbar_\mr{eff}\ra 0$. Then, using the semiclassical calculation we derive $D_{Vp} (t)\propto \hbar_\mr{eff}$ and $C_{Vp} (t)\propto \epsilon^2 e^{2\lambda t}$ (before the Ehrenfest time). Thus, the equivalence holds true for any timescale in the classical limit $\hbar_\mr{eff}\ra 0$ because $D_{Vp} (t)\ll C_{Vp} (t)$ and Eq. (S-\ref{ahoda2}).

\bibliography{../../../../refer_them}